\documentstyle[aps,twocolumn]{revtex} 
 
\begin{document} 
\draft 
\preprint{To be published in Phys Rev B}

\title{Metric tensor as the dynamical variable for 
variable cell-shape molecular dynamics} 

\author{Ivo Souza$^*$ and Jos\'e Luis Martins}  
\address{Departamento de F\'{\i}sica, Instituto Superior T\'ecnico,\\ 
Avenida Rovisco Pais 1, 1096 Lisboa, Portugal\\ 
and\\
Instituto de Engenharia de Sistemas e Computadores,\\
Rua Alves Redol 9, Apartado 13069, 1000 Lisboa, Portugal} 
\date{\today}
\maketitle 
 
\begin{abstract} 
We propose a new variable cell-shape molecular dynamics algorithm
where the dynamical variables associated with the cell are the six 
independent dot products
between  the vectors defining the cell instead of the 
nine cartesian components of those vectors.
Our choice of the metric tensor as the dynamical variable 
automatically eliminates 
the cell orientation from the dynamics. 
Furthermore, choosing for the cell kinetic energy a simple scalar
that is quadratic in the time derivatives of the metric tensor,
makes the dynamics invariant with respect to the choice of the 
simulation cell edges. Choosing the densitary character
of that scalar allows us to have a dynamics that obeys the virial theorem. 
We derive the
equations of motion for the two conditions of constant external pressure 
and constant 
thermodynamic tension. We also show that using the metric as variable
is convenient for structural optimization under those two conditions.
We use simulations
for Ar with Lennard-Jones parameters and for Si with forces and
stresses calculated from first-principles of density
functional theory to illustrate the applications of the method. 
\end{abstract} 
\pacs{PACS: 61.50.Ah, 71.15.Pd, 62.20-x, 62.50+p}

\narrowtext
 
\label{intro}\section{INTRODUCTION} 
 
With the development of new simulation methods and the increase in
available computational
power, molecular dynamics has become an important tool in the simulation
of matter in the condensed state\cite{AT,Frenkel}.
In its earliest applications, molecular dynamics methods were employed to 
simulate systems of interacting particles with a constant density and 
energy, using a simulation cell with a fixed volume and shape and with 
a constant number of particles inside. For extended systems, periodic 
boundary conditions were introduced to reduce finite cell-size effects.

The calculations with constant energy, volume and number of
particles are expected to simulate the thermodynamic properties of the
micro-canonical ensemble. However, in laboratory conditions, 
one often controls
the intensive variables temperature $T$ and pressure $p$, instead of the
extensive variables $E$ and $V$.
Therefore molecular dynamics 
methods were developed to simulate systems at
constant temperature or pressure\cite{AT,Frenkel,Nose,BH,HCA,PR1,PR2}. 
In the case of constant 
pressure simulations, the size and shape of the simulation cell
must be allowed to change. In order to do so, an ``extended system'' is
constructed which includes degrees of freedom for the cell.
A microscopic simulation of the structural, mechanical, and dynamical 
response of material systems to external stress of interest in tribology,
material fatigue and wear, crack propagation, stress induced phase
and structural transformations, lubrication and hidrodynamical phenomena,
is more conveniently done with varying cell shapes.

The dynamics of the cell is fictitious. Therefore there are many
reasonable choices for the equations of motion of the variables
associated with the cell. Traditionally the dynamical
variables were the cartesian components of the vectors defining
the periodicity of the simulation cell. The early choices for
the equations of motion had some invariance problems, and more complicated
equations of motion have been proposed to avoid those problems.

Here we suggest the use of the dot products
between  the vectors defining the simulation 
cell as the variables for the cell
dynamics. We show that using the new variables avoids in
a natural way the problems previously encountered.

\label{newmd}\section{VARIABLE CELL-SHAPE MOLECULAR DYNAMICS}

To simulate a system at constant pressure, one must allow for variations of the
volume and shape of the simulation cell.  
Andersen\cite{HCA} proposed to use the volume $V$ of a cubic
simulation cell as a dynamical variable in an extended hamiltonian, 
thus allowing 
for volume fluctuations driven by the dynamical imbalance between the 
imposed external pressure, $p_{\rm ext}$,
and the actual instantaneous internal pressure, 
$p_{\rm int}$, as given by the virial theorem. 
As the simulation cell is periodically repeated, the
dynamics associated with the cell is fictitious. 
In the extended lagrangian for the dynamics, Andersen included
a fictitious  kinetic energy term associated with the
rate of change of volume,

\begin{equation}
\label{KAnder}K_{\rm cell}^{\rm A}= {W^{\rm A} \over 2} \dot V^2,
\end{equation}

\noindent where $W^{\rm A}$ is a fictitious ``mass'' associated with 
the cell. He also added the term
$U_{\rm cell}=p_{\rm ext}V$,
which is the potential from which the constant external pressure acting
on the cell is derived. 
During the simulations, the volume $V$ fluctuates about an average value 
such that, in the limit of long simulation times,
the time average of the calculated
internal pressure is equal to the chosen external pressure,
$\overline{p}_{\rm int}=p_{\rm ext}$.  
Here we use an overline
to indicate the limit of a time average for long calculation times. 
In those simulations it is the enthalpy $H = E+pV$
that is approximately conserved, not the internal energy, and 
Andersen showed that, assuming ergodicity, his simulation 
method samples the
isoshape-isobaric-isoenthalpic ensemble to an 
accuracy of ${\cal O}\left(N^{-2}\right) $ when 
calculating ensemble averages 
of intensive parameters (${\cal O}\left( N^{-1}\right) $ for extensive 
parameters), where $N$ is the number of particles
in the simulation cell. 
 
Andersen's method is best suited to study equilibrium 
properties of fluids, for
which the shape of the cell is irrelevant. 
To study shear flow (viscosity) in fluids
or to study solids it is not enough to change volume with constant shape.
For example, a given cell shape may be compatible with the
periodicity of one crystal structure and be incompatible with another solid
phase, and so the fixed cell shape may artificially 
prevent the appearance of 
thermodynamically more stable phases. 
In order to study structural 
phase transitions, Parrinello and Rahman\cite{PR1,PR2} extended 
Andersen's method to allow for changes in both the volume and the shape of 
the cell. They used as dynamical variables
the cartesian components
$$h_{ij} = \vec e_i \cdot \vec a_j$$
of the three vectors $\vec a_j$ defining the periodicity of the 
simulation cell. Here $\vec e_i$ are the three orthonormal
vectors that define a cartesian coordinate system.
To generate the dynamics, a fictitious kinetic energy of the cell
$$K_{\rm cell}^{\rm PR} = {W^{\rm PR} \over 2} 
\sum_{i=1}^3 \sum_{j=1}^3 (\dot h_{ij})^2,$$
is included in the lagrangian, where $W^{\rm PR}$ is again a 
fictitious mass.
In the limit of large $N$,
the equipartition principle tells us that the kinetic energy of the 9 variables
of the cell is small compared with the kinetic energy of the $3N-3$ variables
associated with the particles' positions, and the method simulates the
isobaric-isoenthalpic ensemble.

As the kinetic energy of the cell is fictitious, it can be chosen in many
reasonable ways that simulate the same ensemble in the limit of large 
number of particles, $N$, and large simulation times. 
However, different choices of the fictitious cell
kinetic energy yield different dynamics, and one can ask which 
is better or more
convenient. Several authors have pointed out some shortcomings of the
original method of Parrinello and Rahman: it is not invariant 
under modular transformations (defined below),
the consistency between the condition of mechanical equilibrium and the
virial theorem is only verified in the large $N$ limit,
and it has spurious cell rotations\cite{Clev,RW,Ray,LB}.

For a given periodic
system, there are infinite equivalent choices of the basic simulation cell.
If $\vec a_i$ are three vectors commensurate with the periodic system, then
the transformation
$ \vec a_j^\prime  = \sum_k M_{kj} \vec a_k$, with $M$ an integer matrix 
with $|\det M| = 1$, gives another set of vectors describing the periodicity.
It is desirable that the dynamics
should not depend on the particular choice that is made, i.e., the equations
of motion should be formally invariant with respect to the interchange
between equivalent cells (modular transformations)\cite{Clev,RW}.
This characteristic improves the physical content of the simulation, by
eliminating symmetry breaking effects associated with the fictitious part of the
dynamics\cite{RW}. Of course, in the thermodynamic limit ($N\rightarrow\infty$)
these effects vanish, but they may be important in computer simulations, which
may use only a small number of particles. That is often the case in 
first-principles molecular dynamics\cite{CP}.

For long simulation times and constant
applied pressure, the dynamics for the cell should yield 
$\overline{\left({\cal P}_{\rm cart}\right)^{i}_{j}}
  = p_{\rm ext}\delta^{i}_{j}$,
with $\left({\cal P}_{\rm cart}\right)^{i}_{j}$ the internal stress 
in cartesian coordinates given by the stress theorem,
which is a generalization of the virial theorem\cite{NM}. A weaker condition
that is easily checked is that this should
be verified in particular when the cell is restricted to undergo isoshape
fluctuations\cite{Ray}. Andersen's method obeys this condition, while the same
is true for the Parrinello-Rahman dynamics only 
in the large $N$ limit\cite{Clev}.

The orientation in space of the simulation
cell is irrelevant for the structural and thermodynamical description of the
system (principle of material-frame indifference\cite{LB}). However, it is
included in the dynamics if one uses the components of the cell edges as
dynamical variables, and spurious cell rotations have been obtained in actual
simulations with the Parrinello-Rahman method, namely in the simulation of
molecules, whose internal degrees of freedom sometimes cause the internal stress
to be asymmetrical\cite{NK}. These rotations not only are physically irrelevant,
but may complicate the analysis of the simulations' results.
Methods to eliminate them have been proposed, such as 
constraining the matrix of the lattice vectors to be symmetrical\cite{NK} or 
upper triangular\cite{CR} (geometrical constraints), or by symmetrization of the
infinitesimal strain at each time step (dynamical constraint)\cite{LB}. 

\label{metricmd}\section{USING THE METRIC AS A DYNAMICAL VARIABLE} 
  
If $\vec a_1$, $\vec a_2$ and $\vec a_3$ are three linearly independent
vectors
that define the periodic simulation cell and form a right-handed triad,
then all the properties of the 
simulated system depend only on the symmetrical metric tensor,
$$ g_{ij} \equiv \vec a_i \cdot \vec a_j = g_{ji},$$
and not on the orientation of the three vectors in space. 
In our new simulation method, we use the 6 independent components of the metric 
tensor as the dynamical variables for the cell. 
The three diagonal elements 
of the metric give information about the lengths of the lattice vectors, 
and the
three independent off-diagonal elements contain the additional
information about the angles between those vectors. 
The covariant components of the tensor $g$ are related to the matrix
$h \equiv \left( \vec a_1 , \vec a_2 , \vec a_3 \right)$,
the transformation matrix between cartesian and lattice coordinates, by
the relation
\begin{equation} 
\label{metric2}g=h^T h,
\end{equation}
where $h^T$ is the transpose of $h$. 
The one-forms $\vec b^i$ associated with the lattice vectors $\vec a_i$, which
are (except for a factor of $2\pi$) the reciprocal lattice vectors,
are related to the contravariant components of the metric tensor,
$$ g^{ij} \equiv \vec b^i \cdot \vec b^j = g^{ji},$$
and the volume of the unit cell is given by
$$ V=\det h=\sqrt{\det g_{ij}}.$$

The position $\vec r(i)$ of the $i^{\rm th}$ atom in the simulation cell
can be defined by its lattice coordinates $s^k(i)$,
$$\vec r(i)=s^k(i) \vec a_k,$$
where we use the Einstein summation convention for tensorial quantities.
The distance between any two points can be calculated from the metric and
the lattice coordinates, and therefore they completely define the geometry
of the simulation cell.

In the Parrinello-Rahman formalism, the calculation of the total distance
traveled by a particle can be misleading.
Because the unphysical motion due to the rigid rotation of the cell should be discarded,
one cannot in general simply use a two-point formula to calculate, for example,
the mean square displacement of a particle. The correct formula for the distance is
naturally expressed in terms of the metric:

\begin{equation}
\label{distance}\Delta s=\int_{t_0}^{t_1}\sqrt{\dot s^i g_{ij}\dot s^j}dt.
\end{equation}
 
For a fixed cell (i.e., for a fixed $g_{ij}$), the Newton's equations of motion can be 
derived from the lagrangian
\begin{equation}
\label{L1}
\begin{array}{c}
{\cal L}_1\left( s^i(k),\dot s^i(k),g_{ij}\right) = \cr
{1 \over 2} \sum_{k=1}^N m(k) \dot s^i(k)g_{ij}\dot s^j(k)-
U\left(s^i(k),g_{ij}\right),
\end{array} 
\end{equation}
   where the summation is over all $N$ atoms in the cell, 
$m(k)$ is the mass of atom $k$, and the potential energy 
per cell $U$ includes interactions between atoms in different cells.
$U$ is a function of the $3N$ lattice coordinates of the atomic positions and
the 6 independent components of the metric tensor. 
In the examples of a latter section, $U$ is either 
the Born-Oppenheimer energy
from a first-principles pseudopotential local-density calculation
or the potential energy of the Lennard-Jones model.
The momentum canonically conjugate to $s^i(k)$ is
$$ 
\pi _i(k)\equiv {\partial {\cal L}_1 \over \partial \dot s^i(k)}
    =m(k)g_{ij}\dot s^j(k),  
$$ 
and the corresponding hamiltonian is

\begin{equation} 
\label{H1}{\cal H}_1\left( s^i(k),\pi _i(k),g_{ij}\right) =
\sum_{k=1}^N {\pi _i(k)g^{ij}\pi _j(k) \over 2m(k)}
+U\left( s^i(k),g_{ij}\right).
\end{equation}  
 
To construct the extended lagrangian for the cell dynamics, 
we must choose the fictitious kinetic 
energy of the cell, $K_{\rm cell}$, and, for simulations with applied pressure,
add the term $p_{\rm ext}V=p_{\rm ext}\sqrt{\det g_{ij}}$. 
A simple non-negative scalar that is quadratic in the time derivatives of all the
components of $g$ is 
$$
K_{\rm cell}^{0}\left( g_{ij},\dot g_{ij}\right)=
\frac{W^0}{2} \left( \frac {\partial g}{\partial t}\right)_{ji}
\left( \frac {\partial g}{\partial t} \right)^{ij}=
\frac {W^0}{2}\dot g_{ji}\left(g^{ik}\dot g_{kl}g^{lj}\right),  
$$ 
where $W^0$ is a fictitious cell ``mass'' which has the dimensions of 
mass times length squared. The positivity of this term is shown in Appendix A. 
Instead of $K_{\rm cell}^0$, we choose the slightly modified expression,
which is again a scalar, quadratic in $\dot g$, but with a different densitary
character,
$$ 
K_{\rm cell}^{\rm g}\left( g_{ij},\dot g_{ij}\right) =\frac {W^{\rm g}}{2}%
\left( \det g_{ij} \right) \dot g_{ji}\left(g^{ik}\dot g_{kl}g^{lj}\right),
$$ 
where{\bf \ }$W^{\rm g}$ is a fictitious cell ``mass'' with the dimensions of 
mass times length$^{-4}$. 
Alternatively, we may view
${\cal M}^{jikl}\equiv W^{g}\left( \det g_{ij}\right)g^{ik}g^{lj}$ as an
effective ``mass'' tensor.
Although $K_{\rm cell}^{\rm g}$ gives slightly 
more complicated equations of motion for the cell, it has the advantage of 
reducing to Andersen's $K_{\rm cell}^{\rm A}$ (see Eq. \ref{KAnder}) in the
case of isoshape fluctuations of the cell, if we make the identification 
$W^{\rm g}={3 \over 4} W^{\rm A}$, and so the dynamics that it 
generates obeys the virial 
theorem in that limit\cite{Ray}. Since we are using the metric, the orientation
of the cell never appears in the equations. It can also be verified that
$K_{\rm cell}^{\rm g}$ is invariant with respect to modular
transformations of the $\vec a_i$.

The fictitious lagrangian for the extended system in the presence
of an applied external pressure is 
\begin{equation} 
\label{L2} 
\begin{array}{c} 
{\cal L}_2\left( s^i(k),\dot s^i(k),g_{ij},\dot g_{ij}\right) = \frac 
12\sum_km(k)\dot s^i(k)g_{ij}\dot s^j(k)-  \cr
U\left( s^i(k),g_{ij}\right) + 
\frac {W^g}{2}\left( \det g_{ij} \right) \dot g_{ji}g^{ik}\dot 
g_{kl}g^{lj}-p_{\rm ext}\sqrt{\det g_{ij}},   
\end{array} 
\end{equation} 
and the equations of motion for the atomic coordinates are
\begin{equation}
\label{atoms}m(k)\ddot s^i(k)=g^{ij}F_j(k)-m(k)g^{ij}\dot g_{jl}\dot s^l(k),  
\end{equation}
where $F_j(k)\equiv -\partial U/\partial s^j(k)$ are the 
covariant components of the force (which can be viewed as the components in
reciprocal lattice coordinates multiplied by $2 \pi$). 
This equation for the scaled atomic coordinates is identical to the one 
obtained from Parrinello-Rahman's lagrangian, since it does not depend on the 
choice of
$K_{\rm cell}$. It should be stressed that this doesn't imply that the dynamics
of the atoms is the same, because in order to convert from the scaled dynamics to the
actual atomic dynamics we have to use the metric, which is determined by the
cell's dynamics. Hence the importance of a fictitious cell dynamics
which doesn't introduce unphysical symmetry-breaking effects\cite{RW}.

The coupling of the atomic
motion to the cell's motion is made through the second term on the R.H.S. of
Eq. \ref{atoms}, which is independent of the orientation and state of rotation
of the cell; from this, the physical irrelevance of the orientation of the cell
is evident.
 
The equation of motion for the cell variables is derived with the
help of the relation $\frac \partial {\partial g_{kl}}\det g_{ij}
=g^{lk}\det g_{ij}$, giving
 
\begin{equation} 
\label{cell}
\begin{array}{c} 
W^{\rm g}\ddot g_{ij}={1 \over 2\sqrt{\det g_{ij}}}
    \left( { {\cal P}_{ij} \over\sqrt{\det g_{ij}}}
    -p_{\rm ext}g_{ij}\right) + \cr
W^{\rm g} \left( \dot g_{ik}g^{kl}\dot g_{lj}
      -g^{kl}\dot g_{kl}\dot g_{ij}\right) +
{W^{\rm g} \over 2}\left( \dot g_{kl}g^{lm}\dot g_{mn}g^{nk}\right) g_{ij}, 
\end{array} 
\end{equation} 
where the contravariant components of the internal stress are (see Appendix B)
\begin{equation} 
\label{stress2}{\cal P}^{ij}=\sum_km(k)\dot s^i(k)\dot s^j(k)-2%
\frac{\partial U}{\partial g_{ij}}. 
\end{equation} 
The instantaneous internal pressure averaged over the cell is
${1 \over 3V} {\rm Tr}{\cal P}_j^i$, and it can also be obtained from
${\cal H}_1$\cite{Ray} or ${\cal L}_1$: 
$$ 
p_{\rm int}=-\left( \frac{\partial {\cal H}_1}{\partial V}%
\right) _{\pi _i(k)}=\left( \frac{\partial {\cal L}_1}{\partial V}\right) 
_{\dot s^i(k)}.  
$$ 
 
Defining the  momentum canonically conjugate to the metric tensor 
 $$ 
\Pi ^{ij}\equiv \frac{\partial {\cal L}_2}{\partial \dot g_{ij}}%
=W^{\rm g}\left( \det g_{ij}\right) g^{ik}\dot g_{kl}g^{lj}=\Pi ^{ji},  
$$ 
 
\noindent the conserved extended hamiltonian can be written as 
 
\begin{equation} 
\begin{array}{c} 
\label{H2}{\cal H}_2\left( s^i(k),g_{ij},\pi _i(k),\Pi ^{ij}\right) =\sum_k%
\frac{\pi _i(k)\pi ^i(k)}{2m(k)}+ \cr
U\left( s^i(k),g_{ij}\right) 
+\frac{{\Pi^i}_k {\Pi^k}_i}{2W^{\rm g}\det g_{ij}}+p_{\rm ext}V.  
\end{array} 
\end{equation} 

In the following sections it will be convenient to define a 
symmetrical covariant internal stress tensor as
$$\sigma^{ij}_{\rm int} = - 2  \frac{\partial U}{\partial g_{ij}},$$
which contains the contributions from the potential energy $U$ to $\cal P$. 
 
\label{anisotropicstress}\section{ANISOTROPIC EXTERNAL STRESS} 
 
A constant applied anisotropic stress is in general non-conservative, and thus 
there is no conserved extended hamiltonian in a constant anisotropic stress 
simulation\cite{Clev,RL}. 
Of course some experimental situations are essentially non-conservative,
and therefore best simulated by an appropriate non-conservative 
dynamics\cite{Clev,RL}. In this section we will present a conservative 
dynamics, but one should keep in mind that the simulation should be taylored
to the problem.

Molecular dynamics simulations with an applied
anisotropic stress were first proposed by Parrinello and Rahman\cite{PR2}. 
Ray and Rahman\cite{RR} later showed that the original formulation was valid 
only in the limit of small deformations, and they proposed an extension 
valid for finite deformations, in which it is the thermodynamic tension
(defined below),
not the stress, that is kept constant, and the quantity 
that is approximately conserved during the simulation is the generalized 
enthalpy of Thurston\cite{Thur}. 
This approach is based on the fact that, if the external stress is allowed 
to change when the cell deforms, so as to keep the thermodynamic tension 
constant, the virtual work of the stress upon deformations of the cell is 
conservative, and so that stress is derivable from a potential, which can be 
used to construct an extended hamiltonian. The thermodynamic tension is given 
by\cite{RR,Thur} 
\begin{equation}
\label{thermotension} 
\tau =\frac V{V_0}h_0h^{-1}\sigma_{\rm ext}^{\rm cart}\left( 
h^T\right) ^{-1}h_0^T,  
\end{equation} 
where $h_0$ and $V_0$ are the reference lattice and its volume,
and $\sigma_{\rm ext}^{\rm cart}$ is the external stress
in cartesian coordinates. For $h=h_0$, $\tau$ and $\sigma_{\rm ext}^{\rm cart}$ 
coincide. The virtual work $\delta W$ done by an external stress on 
the faces of the cell during an infinitesimal deformation of the cell in the
state $h$ is\cite{RL,RR},  
 
\begin{equation} 
\label{dW1}\delta W=V_0Tr \left(\tau \delta \varepsilon\right) ,  
\end{equation}
 
\noindent where $\varepsilon$ is the strain tensor for the lattice $h$ measured
from the reference lattice $h_0$. We see that $V_{0} \tau$ is the
thermodynamic variable conjugate to the strain. Thus, for fixed $\tau$, the
differential is exact, and so we can 
integrate $\delta W$ over a finite deformation, to obtain the elastic energy 
 
$$ 
U_{\rm cell}\left( h \right)=\int_{h_0}^h\delta W=V_0Tr(\tau \varepsilon ).
$$ 
 
The generalized enthalpy of Thurston is given by\cite{Thur} 
$$ 
\tilde H \equiv E+V_0Tr \left( \tau \varepsilon \right),  
$$ 
where $E$ is the energy of the system. For our metric-based formulation,
it is desirable to use the metric, instead of the strain, as the thermodynamic variable.
In order to find what is the conjugate variable, we have to
express $\delta W$ in terms of infinitesimal variations of the metric tensor. This can 
be done for a symmetrical (i.e., torque-free) external stress, which does no
work in pure rotations of the cell. The result is given in Eq. (3.5) of Ref\cite{RR},
and, expressed in tensorial notation is a simple expression,
$$ 
\delta W=\frac 12Tr\left( \sigma_{\rm ext}^{ij}\delta g_{jk}\right).
$$  
The thermodynamic variable conjugate to the metric is therefore
the external stress in
contravariant lattice coordinates. Keeping $\sigma_{\rm ext}^{ij}=\sigma_{\rm ext}^{ji}$
constant when the cell deforms
thus leads to a conservative external stress, derived from the potential 
\begin{equation} 
\label{Ucell2}U_{\rm cell} \left( g \right)=\frac 12\sigma_{\rm ext}^{ij}%
\left( g_{ij}-g_{0_{ij}}\right),  
\end{equation} 
\noindent where $g_0=h_{0}^{T}h_0$ is some reference metric.
Since $\sigma^{ij}_{\rm ext}$ is fixed, one can drop the constant term
$-\frac 12 \sigma^{ij}_{\rm ext}g_{0_{ij}}$ from the definition of $U_{\rm cell}$,
obtaining
\begin{equation} 
\label{Ucell3}U_{\rm cell} \left( g \right)=\frac 12\sigma^{ji}_{\rm ext}g_{ij},
\end{equation}
which is independent of a reference configuration and quite compact
when compared with the definition of $\tau$ in Eq.~\ref{thermotension}.

The condition that
$\sigma_{\rm ext}^{ij}$ is constant is equivalent to requiring
$\tau $ to be constant because its cartesian coordinates are
$$ 
\tau _{ab}=\frac 1{V_0}h_{0_{ai}}\sigma ^{ij}_{\rm ext}h_{0_{jb}}^T,  
$$ 
as can be seen using Eq. \ref{tensord} from Appendix B\cite{RR_did_it}. Nevertheless,
to a given $\sigma^{ij}_{\rm ext}$ doesn't correspond a unique thermodynamic
tension, because $h_0$ is arbitrary. All the physical
information is contained in $\sigma^{ij}_{\rm ext}$ and $g_{ij}$, except for the (arbitrary)
choice of axes. The thermodynamic tension fixes the choice of axes and also a
reference state, through $h_0$.
Notice that from
the transformation law for the contravariant components of the stress,
$\sigma^{'}_{\rm ext} = {\det h^{'} \over \det h} \left[ {h^{'}}^{-1} h \sigma_{\rm ext} h^{T} \left( {h^{'}}^{T}
 \right) ^{-1} \right]$ 
obtained from Eq. \ref{tensord}
keeping $\sigma^{\rm cart}_{\rm ext}$ constant, 
we can conclude that $U_{\rm cell}$ as given by
Eq. \ref{Ucell3} is invariant under modular transformations\cite{Clev}.

To grasp the physical meaning of $\sigma_{\rm ext}^{ij}$, let's consider the force acting
on the face of the cell opposing edge $i$.
Using Eqs. \ref{Ucell3} and \ref{metric2}, we obtain (in cartesian coordinates)
\begin{equation}
\label{Fcell}\vec {\cal F}^i \equiv - \frac {\partial U_{cell}}{\partial {\vec a_i}} =
- \sigma^{ij}_{\rm ext} \vec a_j,
\end{equation}
showing how the force on the face $i$ is related to the stress.

The new extended lagrangian can be obtained from ${\cal L}_2$ given by 
Eq. \ref{L2}, by replacing $p_{\rm ext}V$ by the new $U_{\rm cell}$,
Eq. \ref{Ucell3}. The equation of motion for the atoms, Eq. \ref{atoms}, 
remains unchanged, and the equation of motion for the cell is obtained from 
Eq. \ref{cell} by replacing $p_{\rm ext}g_{ij}$ by
$\frac 1V \sigma^{\rm ext}_{ij}$:
\begin{equation} 
\label{cell_aniso}
\begin{array}{c} 
W^{\rm g}\ddot g_{ij}={1 \over 2 \det g_{ij}}
    \left( {\cal P}_{ij} 
    -\sigma^{\rm ext}_{ij}\right) +
W^{\rm g} \left( \dot g_{ik}g^{kl}\dot g_{lj}
      -g^{kl}\dot g_{kl}\dot g_{ij}\right) + \cr
{W^{\rm g} \over 2}\left( \dot g_{kl}g^{lm}\dot g_{mn}g^{nk}\right) g_{ij}.
\end{array} 
\end{equation} 
The conserved hamiltonian is
\begin{equation} 
\label{Haniso}
\begin{array}{c} 
{\cal H}_{\rm aniso}\left( s^i(k),g_{ij},\pi_i(k),\Pi^{ij}\right) =\sum_k 
\frac{\pi _i(k)\pi ^i(k)}{2m(k)}+ \cr
U\left( s^i(k),g_{ij}\right) 
+\frac{{\Pi^i}_k {\Pi^k}_i}{2W^{\rm g}\det g_{ij}}+
\frac 12\sigma^{ji}_{\rm ext}g_{ij}.  
\end{array} 
\end{equation} 

In specific applications, it may be desirable to impose a constant external 
pressure, $p_{\rm ext}$, plus a constant thermodynamic tension.
Note that the stress tensor associated with a
constant pressure is $\sigma_{\rm ext}^{ij}=p_{\rm ext}V g^{ij}$, 
and so constant pressure is not a particular case of constant
thermodynamic tension. The generalization is straightforward, and in this case,
when considering only isoshape fluctuations of the cell, the equation of motion
for the cell becomes 
 
$$ 
W_{\rm A}\ddot V\delta _j^i=\frac 1V {\cal P}_j^i-\left( \frac 1V {\sigma_{\rm ext}}_j^i+
p_{\rm ext}\delta_j^i\right),  
$$ 
 
\noindent where $W_{\rm A}$ is Andersen's cell mass. This equation shows that the off-diagonal
elements of $%
\left( {\cal P}-\sigma_{\rm ext} \right) _j^i$ are restricted 
to be zero, and the diagonal elements are restricted to take equal values, 
at all times: by imposing a fixed cell shape, we have arrived at an 
isotropic total stress, as should be expected on physical grounds. In 
equilibrium $\overline{\ddot V}=0$, and so the average of 
each diagonal component of $\frac 1V{\cal P}_j^i$ equals 
$\frac 1V\left({\sigma}_{\rm ext}\right)_j^i + p_{\rm ext}  {\delta}_j^i$, which implies
$\overline{p_{\rm int}}=\overline{p_{\rm ext}}+
\overline{\frac {1}{3V}\rm{Tr} \left({\sigma_{\rm ext}}\right)^i_j}$,
where the R.H.S. is the total external pressure.
This shows that our method obeys the virial theorem in the case of isoshape fluctuations
of the cell (the proof in Ref.\cite{Ray} mentioned in Section \ref{anisotropicstress}
was for applied pressure only).

\label{optimization}\section{STRUCTURAL OPTIMIZATION}

A problem encountered in the simulation of materials is the 
determination of the
equilibrium structure of a crystal at a given pressure (or anisotropic 
stress) predicted by a given model $U\left(s^i(k),g_{ij}\right)$
of its total energy. This can, in principle, be achieved by the minimization (under
the appropriate constraint) of $U$, which is quite difficult because it is a
multivalleyed function of many variables. A practical strategy is to
use a simulated annealing to bring the configuration to a deep valley,
followed by a search of a minimum in that valley. The annealing step
can be carried out by the variable cell shape molecular dynamics
described previously coupled to a thermostat, brownian dynamics forces,
or a periodic rescaling of the velocities. The local minimization
can be done efficiently if one has the gradient of the function
to be minimized.

If we want to obtain the crystal structure at zero temperature and for an
applied pressure of $p_{\rm ext}$, we must minimize its enthalpy,
$$
H \left(s^i(k),g_{ij}\right) = U \left(s^i(k),g_{ij}\right)
+ p_{\rm ext} \sqrt{\det g_{ij}}.
$$
The gradient of the enthalpy with respect to atomic positions is
$$
{\partial H \over \partial s^i(k)} = {\partial U \over \partial s^i(k)}
= -F_i(k),
$$
which is minus the covariant components of the force on that atom.
Notice that in molecular dynamics it is the contravariant components,
$F^i(k)=g^{ij}F_j(k)$ that appear in the equation of motion.
The gradient of the enthalpy with respect to the metric is
$$
{\partial H \over \partial g_{ij}} = {\partial U \over \partial g_{ij}}
+ p_{\rm ext} {\partial \over \partial g_{ij}} \sqrt{\det g_{ij}} =
-{1 \over 2} \sigma_{\rm int}^{ij} + 
{1 \over 2} p_{\rm ext} g^{ij} \sqrt{\det g_{ij}}.
$$
The minimum is obtained when the forces are zero and when the mixed stress 
tensor divided by the volume is the pressure times the identity tensor,
$$
{1 \over \sqrt{\det g_{ij}}} {\sigma_{\rm int}}^i_j = p_{\rm ext} \delta^i_j
$$
as desired.

If we want to obtain the crystal structure for a fixed thermodynamic tension, 
then we must minimize the generalized enthalpy,
\begin{equation} 
\label{gen_enth}
\tilde H \left(s^i(k),g_{ij}\right) = U \left(s^i(k),g_{ij}\right)
+{1 \over 2} \sigma_{\rm ext}^{ij} g_{ij}.
\end{equation}
The gradient of the generalized enthalpy with respect to the atomic lattice
coordinates is still minus the covariant force on the atoms, and the gradient with
respect to the metric is
$$
{\partial \tilde H \over \partial g_{ij}} = {\partial U \over \partial g_{ij}}
+  {1 \over 2} \sigma_{\rm ext}^{ij}=
-{1 \over 2} \sigma_{\rm int}^{ij} + {1 \over 2} \sigma_{\rm ext}^{ij},
$$
which is zero when the internal stress is equal to the desired applied
stress, $\sigma_{\rm int}^{ij} = \sigma_{\rm ext}^{ij}$.

\label{applications}\section{APPLICATIONS}

In this section, we apply the method to the study of structural phase transitions
and structural optimization.
The isobaric-isoenthalpic ensemble, besides being somewhat unusual\cite{RGH},
is not the most adequate to study transitions induced by pressure or stress,
because it does not allow for the exchange of heat with the surroundings. 
There are several methods described in the literature to perform
simulations at constant temperature by connecting the system to
a ``heat bath''\cite{Nose,BH}. In our examples we use Langevin 
molecular dynamics\cite{BH}, but first we checked that 
in our simulations the generalized enthalpy was conserved
in the absence of the heat bath.
In the course of a simulation of a structural transformation
the release of heat of transformation must be dissipated to the
heat bath, this takes some time and therefore the temperature of
the system may rise to values quite above those of the heat bath.

As our first example we simulated a silicon crystal under a
constant pressure using first-principles molecular dynamics.
Pioneering examples of first-principles molecular dynamics with 
variable cell shape include the optimization of the structure parameters of 
MgSiO$_3$ under pressure\cite{WMP} and the
structural transition of silicon under pressure\cite{FCBTP}.
In the first case the dynamical variable for the cell was the strain tensor,
in the second the lattice vectors.

In our simulation of Si, the energy, forces and stresses were 
calculated within the local-density 
approximation, using a pseudopotential\cite{TM} and a plane-wave basis set with a cutoff
of 16~Ry\cite{Pickett}.
The simulation cell contained 8 atoms, initially disposed in a
diamond structure, with lattice constant $a=9.435$~a.u. 
The applied pressure was 25 GPa. The equations of motion
were integrated with a Beeman algorithm\cite{Beem}.
The time step was $h=200$~a.u., and the cell
``mass'' $W^{g}=10$~a.u. Langevin dynamics with a
viscosity damping constant of $\gamma = \frac{2}{m\left(Si\right)}$,
where $m\left(Si\right)$ is
the atomic mass of silicon, was used to simulate a heat bath
with a temperature of 300K.

It is well-known that silicon undergoes several phase transformations 
with increasing pressure, and its pressure-volume phase diagram
has been extensively studied\cite{HMMS}.
Starting from a diamond lattice, the structure
changes at $\sim$11 GPa into $\beta$-Sn, and between 13 and 16 GPa 
transforms into simple
hexagonal. Other densely packed phases appear at around 38 GPa.
In the first $\sim 0.7$~ps (200 steps) of the simulation, we observed (Fig.~1)
that the volume of the simulation cell
was fluctuating around a value that corresponds to the volume of the 
metastable diamond structure of Si at 25 GPa ($V\sim 885$ a.u.\ for the
8 atoms of the conventional cubic unit cell). There was 
then a rapid drop in the volume,
accompanied by a rapid rise in the ionic
temperature to around 3500K (well above the melting point).
The simulation was interrupted after 1000 steps, well before equilibrium
with the thermal bath was reached. 
After the transition, the volume of the cell oscillated
around 650~a.u., slightly below the volume of the stable
simple hexagonal structure at that pressure, but above the density
of the close-packed structures. Remembering that at atmospheric pressure
Si contracts upon melting, and considering
the high temperatures of the simulation, our results indicate that at 
high pressures, the liquid phase may still be denser than the solid phase.

For the purpose of illustrating a molecular dynamics
method the origin of the forces is irrelevant,
therefore we used a Lennard-Jones model,
with the constants adjusted to simulate argon, for the  other examples
in this article, 
as the computational demands are much lower.
In our second example, we started with a cubic simulation cell with 32 argon atoms
in an fcc lattice, and increased one diagonal contravariant component of the
external stress, $\sigma_{\rm ext}^{33}$, 
linearly with time 
from zero to $15 \times 10^{-5}$ a.u.\ 
for the first 4000 simulation steps, and held thereafter
$\sigma_{\rm ext}^{33}$ at that
value. All other applied stress components were kept at zero.
This corresponds to a situation of uniaxial compression.
During the first 10000 steps of the simulation the system is
kept in contact with a heat bath at 10~K, at which point we minimize
the generalized enthalpy (Eq.\ \ref{gen_enth}) using a method by 
Davidon\cite{Dav}. The minimization is 
obtained in 96 steps which is approximately the number of variables, 
indicating that the heat bath kept the system near a quadratic region
of the potential.

During the compression the system yields for an applied stress
of $\sim~0.1$~GPa (after $\sim 2500$ steps), and due to the rearrangement of
the atoms, the applied stress drops to a minimum of $\sim 0.07$~GPa and then rises
again gradually, as the thermodynamic tension is increased.
After the structural rearrangement,
the argon is still in a distorted fcc lattice, but the stress is now applied in
a [110] direction instead of the initial [100] direction, and the area on
which the force is applied is $\sim \sqrt 2$ times larger. 
The yield was accompanied by a rapid rise in the ionic temperature up to
$\sim 33$~K. The heat was gradually dissipated, and at around step 4000 the
temperature was back to 10~K. 

Figure~\ref{fig_stress} shows the evolution of two of the contravariant
lattice components of the internal stress, $\sigma_{\rm ext}^{33}$ and
$\sigma_{\rm ext}^{31}$,
compared with the corresponding imposed external stress
components. At first the internal stress oscillates around the 
external values, in particular it accompanies the rise in applied stress.
When the system yields we observe a dramatic increase in the amplitude
of the stress oscillations, which are then damped with time. Finally, in the
minimization step the internal and external stress are identical within the
precision demanded in the minimization ($10^{-5}$).

The contravariant components of the stress tensor are not what we are used
to call stress (their dimensionality is energy per area),
so we show in Figure~\ref{fig_stress_cart} the evolution of the
cartesian components $\sigma_{\rm zz}$ of the applied and internal stress,
where we chose the $z$ axis to be in the direction of the applied stress.
One can see that the cartesian components of the applied stress are 
not constant
when the contravariant components of the stress are constant, and that the
oscillations of the internal stress are magnified, but they track each other,
and they are identical at the end of the enthalpy minimization, as desired.

The yield is also apparent in the plot of the potential component
of the generalized enthalpy (Eq.~\ref{gen_enth})
as a function of time (Fig.~\ref{fig_enth}).
First we observe an increase of the enthalpy during load due to the
work done on the system by the uniaxial stress. When the system
yields there is a
strong decrease of the potential component of the enthalpy
even while we continue loading the system, showing 
that energy is transfered to the kinetic components, and later dissipated
to the heat bath. Only near the
end of the loading cycle do we see the enthalpy rising again.
During the annealing steps there is a rapid initial decrease of the
enthalpy, meaning that the minimization procedure rapidly reaches
the valley of the multi-variable function, but then takes
some time to reach the minimum.

The best way to observe the yield is from the plot of the 
evolution of the lattice constants (Fig.~\ref{fig_latt}). From 
the initial slopes one could extract the Young and Poisson moduli for the
system.
After the yield we see that the three lattice constants are different
from each other,
and that they are rapidly determined by the minimization procedure.
At the end of the simulation and after the inspection of the angles
we obtain a monoclinic simulation cell, which is in reality a supercell
of the orthorhombic system one should
expect when loading an fcc crystal in the [110] direction.

A movie of the simulation shows that the
$\lbrace 100 \rbrace$ planes parallel to the uniaxial stress become
distorted close-packed planes by compression along the direction of the stress
and expansion along the perpendicular direction. 
A similar simulation was performed by Ray and Rahman in Ref.\cite{RR2}.
They found an fcc to close-packed transition, with the final
structure presenting stacking-faults.
 
Our final example is of a structural optimization under pressure.
We start from conditions quite away from equilibrium, perform
2000 steps in contact with an heat bath, and then switch to a gradient
minimization. Our target pressure is 0.3~GPa and we simulate 16 argon atoms
with a Lennard-Jones potential. The final structure is close-packed 
and corresponds to a stacking of close-packed planes that is neither
fcc nor hcp. During part of the simulation the temperature is well
above melting, so the memory of the initial configuration is lost.
The evolution of the potential contribution to the enthalpy is shown
in Fig.~\ref{fig_iso}. In the inset of that figure that magnifies 
the minimization part of the simulation, one can see that we obtain the
enthalpy of Lennard-Jones argon at that pressure.  The true
minimal structure is not reached because the energy cost of
the stacking faults is too small, so the procedure only finds a 
deep local minimum.

In principle the calculation of the energies, forces and stresses
can be carried out within the metric formalism, and therefore one never needs
to construct the lattice vectors, that is the matrix $h$. Our code for the
Lennard-Jones interaction was written to test the present formalism
and is fully implemented in the metric language. It never uses the matrix $h$.
Our pseudopotential plane-wave code is based on Sverre Froyen's
Berkeley code, which stored atomic positions and $k$-vectors in lattice
and reciprocal lattice coordinates respectively, that is in contravariant
and covariant coordinates. In fact the stress was calculated by applying the
chain rule 
$\partial U/ \partial h_{ij} = 
(\partial U/ \partial g_{kl})( \partial g_{kl}/ \partial h_{ij})$,
so it was easy to convert the program to the present formalism.
The calculations involving the separable non-local pseudopotential
projectors are easier to perform 
in cartesian coordinates, so for that specific case we
construct from the metric $g$ a triangular $h$ and proceed in cartesian
coordinates. The arbitrary choice of the orientation of $h$ has, of course, no
effect in the results of the calculation. Our plane-wave
code also has a old, but convenient, symmetry recognition package that
only works for the conventional orientation of the unit cell. 
If one wants to perform simulations with fixed symmetry, than one has to put
``by hand'' the desired orientation of $h$ before using the package.
Replacing those two parts of the code to avoid using the matrix $h$ is a 
straightforward, but tedious job, it is much easier to use the tested old
subroutines and construct a matrix $h$ whenever it is needed.

\label{conclusions}\section{CONCLUSIONS}

We have shown that the metric is a very convenient dynamical
variable to use in molecular dynamics simulations with variable cell-shape.
As the cell part of the dynamics is fictitious, there is no unique choice 
of the kinetic energy.
to be included in a lagrangian or hamiltonian formulation.
The use of the tensorial notation in a metric formalism, with the requirement
that the energy functions must be scalars, restrict our choice of
those functions. The simplest expression for the cell kinetic energy
has several properties that were not present
in early expressions, namely absence of rigid rotations
and invariance with respect to modular transformations.
With a convenient choice of the
densitary character of the kinetic energy, the virial theorem is
also satisfied for isoshape fluctuations.
For anisotropic stress, the simplicity of Eq.~\ref{Ucell3}
contrasts with the definition of thermodynamic tension,
(Eq. \ref{thermotension}) and its dependence on a reference
cell.

  From our kinetic and potential functions
for the cell metric, we derived the equations of motion for 
variable cell shape molecular dynamics
under the conditions of constant applied pressure and anisotropic
applied thermodynamic stress. 
We also showed that the optimization of structures under both conditions
can be naturally expressed in the metric language.
Simulations of silicon with first-principles forces and
argon with empirical Lennard-Jones forces were used to 
illustrate the applications of our equations of motion and minimization
procedures to the
study of systems under applied pressure or stress.

\section{ACKNOWLEDGMENTS}

This work was supported by
PRAXIS XXI grants 2/2.1/FIS/467/94, 2/2.1/FIS/26/94 and BD5037/95,
and by the US Department of Energy Grant No. DEFG02-91-ER45439.

\section{APPENDIX A}

To prove that $K_{\rm cell}^g$ is non-negative, we have to show
that $\dot g_{ji}g^{ik}\dot g_{kl}g^{lj}=Tr\left( \dot gg^{-1}\dot gg^{-1}\right)$
is non-negative. Using Eq. \ref{metric2} and the usual properties of the trace,
we find that

$$
Tr\left( \dot gg^{-1}\dot gg^{-1}\right)=2Tr\left[ \left( X+X^T\right)X\right],
$$

\noindent where $X\equiv \dot hh^{-1}$. Writing the rightmost $X$ as
$\frac 12 \left( X+X^T\right)+
\frac 12\left(X-X^{T}\right)$ and using $Tr \left( X^{T}X\right)\ge 0$, we arrive
at the desired result. It will be useful to derive this result in the hamiltonian
formalism:

Defining $H_{ba,ij}\equiv \frac {h_{bi}h_{aj}}{\det h}$, where some $h$ compatible
with $g$ was chosen, and defining

$$
G_{kl,ij}\equiv\sum_{ab} H_{kl,ab}^{T}H_{ab,ij}=G_{ij,kl},
$$

\noindent where $H_{kl,ab}^{T}\equiv H_{ab,kl}$, we can write the
kinetic energy of the cell as

\begin{equation}
\label{Kcell2b}K_{cell}^g=\frac{\Pi^{ij}G_{ij,kl}\Pi^{kl}}{2W^g}=
\frac {\sum_{ab} P_{ab}^{2}}{2W^g},
\end{equation}

\noindent where $P_{ab}\equiv H_{ab,kl}\Pi^{kl}=P_{ba}$ is a new generalized momentum
for the cell. The canonically conjugate coordinate is
$Q^{ab}\equiv \left( H_{ab,ij}^T\right)^{-1}g_{ij}$, as can be seen using the
Poisson brackets relations between canonically conjugate variables
(in this last expression, $H^T$ is viewed as a 9x9 matrix with indices $ab$ and
$ij$). The relation between the variables (Q,P) and the variables $(g,\Pi)$ is
similar
to the relation between $\left( \overrightarrow{r},\overrightarrow{p} \right)$ and
$\left( \overrightarrow{s},\overrightarrow{\pi} \right)$.

  From Eq. \ref{Kcell2b}, it is clear that $K_{\rm cell}^g$ is positive and
contributes with 6 distinct quadratic terms to the energy, and so the
equipartition theorem applies to the degrees of freedom of the cell when they
are in contact with a heat bath.

\section{APPENDIX B} 
 
The stress is not a true tensor, but a tensorial density\cite{Bri}, thus
transforming differently from tensors under a change of coordinates whose 
jacobian is not unity. The transformation of a tensorial density ${\cal D}^{ij}$
from cartesian to lattice coordinates is given by\cite{Bri} 
 
\begin{equation} 
\label{tensord}{\cal D}^{kl}=\left( \det h\right) 
h_{ki}^{-1}{\cal D}_{\rm cart}^{ij}\left( h^T\right) _{jl}^{-1}. 
\end{equation} 
 
The average symmetrized internal stress in cartesian coordinates is obtained
from the stress theorem\cite{NM}: 
 
\begin{equation} 
\label{stresst}
\begin{array}{c} 
{\cal P}_{\rm cart}^{ij}=-\frac 1V\left( \frac { \partial E}{\partial \varepsilon_{ij}%
^{'}} \right) _{\varepsilon ^{'}=0}  = \cr%
\frac 1V\left[ \sum_am(a)v^i(a)v^j(a)-\left( \frac{%
\partial U}{\partial \varepsilon _{ij}^{'}}\right) _{\varepsilon ^{'}=0}\right] ,  
\end{array} 
\end{equation} 
 
\noindent where $E$ is the internal energy, $v^{i}(a)$ is the velocity of the atom $a$,
and $\varepsilon ^{'}$ is the (symmetrical) Lagrangian strain corresponding to
a rotation-free infinitesimal homogeneous deformation
given by $h^{'}=\left( 1+ \varepsilon ^{'} \right) h$\cite{footnote1},
from the state $g$ to the state $g^{'}$.
In order to convert to lattice coordinates, we use Eq. \ref{tensord} and apply the
chain rule $\frac{%
\partial E}{\partial \varepsilon _{ij}^{'}}=\frac{\partial E}{\partial 
g_{kl}^{'}}\frac{\partial g_{kl}^{'}}{\partial \varepsilon _{ij}^{'}}$ together 
with the relation $g^{'}-g=2h^T\varepsilon ^{'}h$\cite{PR2}. The result is

\begin{equation}
\label{virial1}{\cal P}^{ij}=-2\frac{\partial E}{\partial g_{ij}}=%
-2\frac{\partial K}{\partial g_{ij}}-2\frac{\partial U}{\partial g_{ij}},
\end{equation}

\noindent where $K$ and $U$ are the kinetic and the potential energy of the atoms in the
cell, respectively, and the factors of two arise from formally treating
$g_{ij}$ and $g_{ji}$ as independent variables. The kinetic energy
is given by the first term in the R.H.S. of Eqs. \ref{L1} or
\ref{H1}. Using these expressions, we find, with the help of the relation
$\frac{\partial g^{nl}}{\partial g_{km}}=-g^{nk}g^{ml}$, that

$$
\left( \frac{\partial K}{\partial%
g_{kl}}\right) _{\pi_{i}\left( k\right)}=-\left( \frac{\partial K}%
{\partial g_{kl}}\right) _{\dot s^{i}\left( k\right)},
$$

\noindent and so it must be made clear whether it is the $\{\pi_{i}\left( k\right)\}$ or
the $\{\dot s^{i}\left( k\right)\}$ that are kept constant when taking the derivatives in
Eq. \ref{virial1}. It is easily seen that, in order to obtain the correct
kinetic internal stress (see Eq. \ref{stresst}) when
transforming Eq. \ref{virial1} back to cartesian coordinates, the $\{\pi_{i}(k)\}$ must
be kept constant, and thus we conclude that

$$
{\cal P}^{ij}=-2\left( \frac{ \partial {\cal H}_1}{\partial g_{ij}}\right)_%
{\pi_{m}(k)}=2\left( \frac{\partial {\cal L}_1}{\partial g_{ij}}\right) _{\dot s^{m}(k)},
$$

\noindent which is Eq. \ref{stress2}.
 
We note that, because $\frac 1{\sqrt{\det g_{ij}}}$ is a scalar 
capacity\cite{Bri}, and ${\cal P}_{ij}$ is a tensorial density, the  
product $\frac 1{\sqrt{\det g_{ij}}%
}{\cal P}_{ij}$ that appears in Eq. \ref{cell} is a true tensor, and thus has the
same densitary character as the pressure term in the same equation, 
$p_{\rm ext}g_{ij}$.

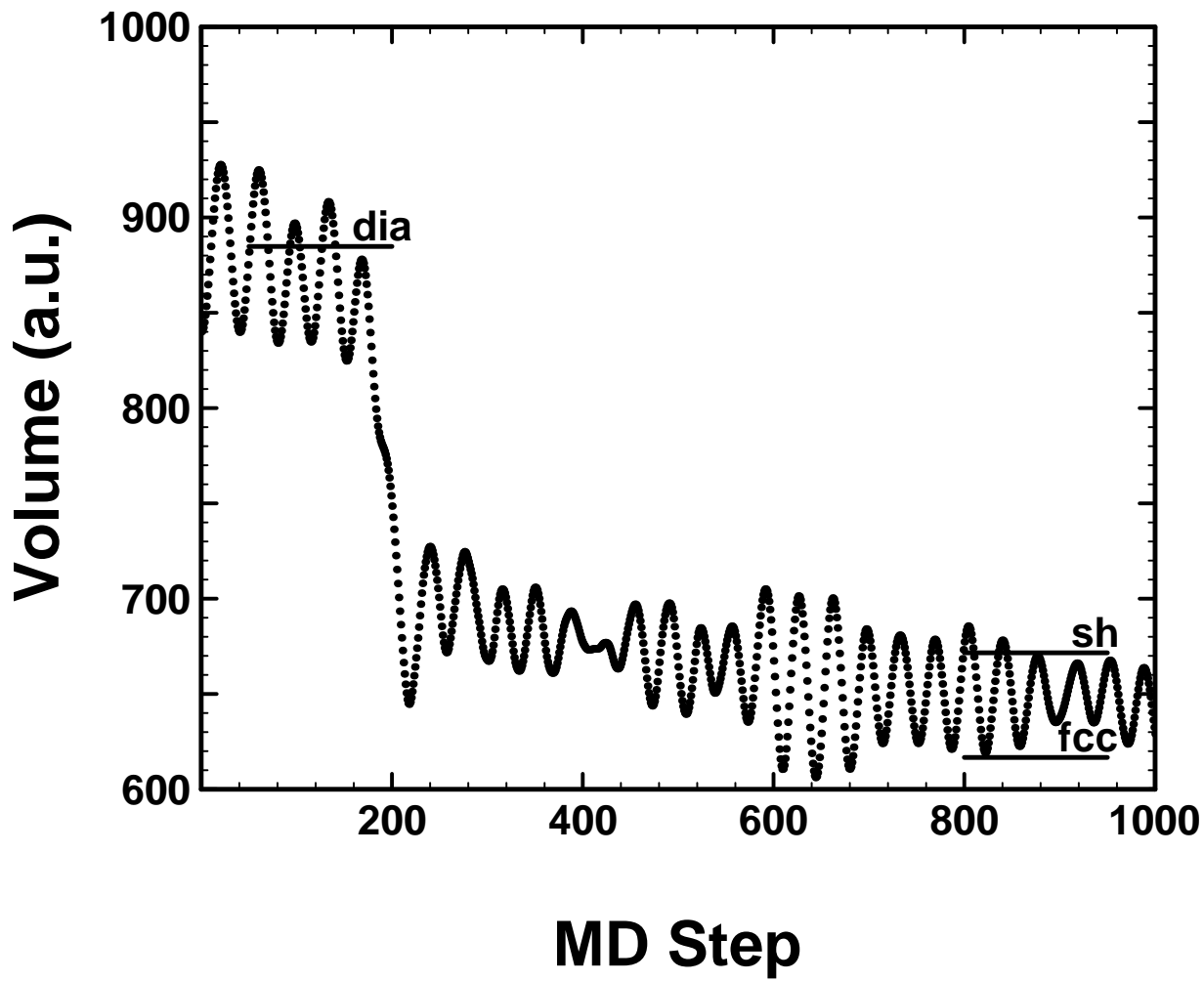
\begin{figure}   
\caption{
The volume (in atomic units) of an eight-atom Si cell is shown as a function of the step
of a first-principles molecular dynamics with an applied pressure of 25~GPa.
The dots indicate the simulated data and the three horizontal lines
indicate the volumes of the diamond, simple hexagonal and face centered cubic
structures of eight Si atoms at a 25~GPa pressure. At that pressure
the stable phase is the eight-fold coordinated simple hexagonal.
The volume starts by oscillating around the volume of the initial
diamond phase, but after 200 steps shows a rapid decrease to values near
the equilibrium value at 25~GPa. The heat released during that 
transformation melts the system and at the end of the short simulation
it has not yet reached equilibrium with the surrounding heat bath.
The simulation seems to indicate that at a pressure of 25~GPa, like at
zero pressure, Si contracts upon melting.
}
\label{fig_si}
\end{figure}

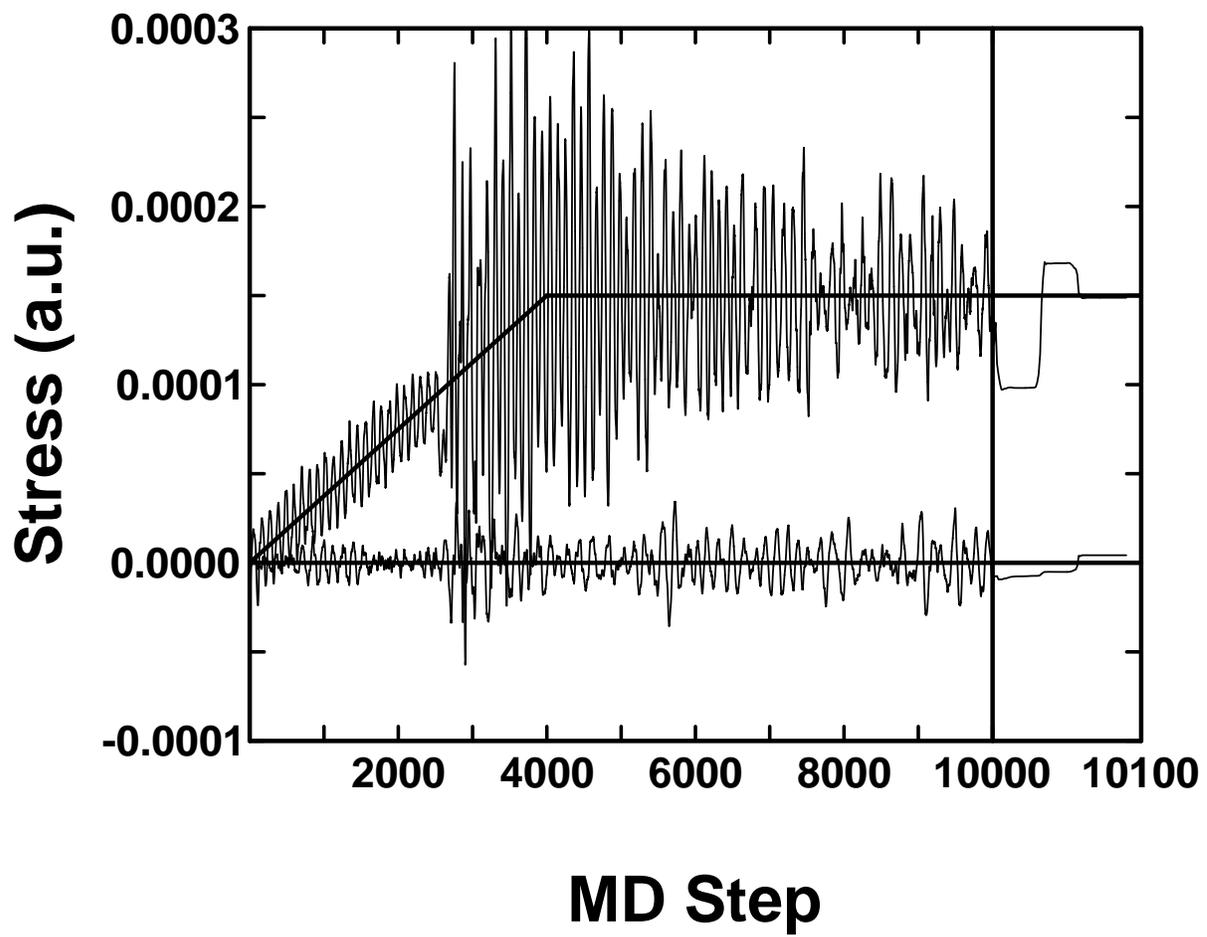
\begin{figure}   
\caption{
Two of the contravariant (lattice) components of the applied and internal
stress tensors (in atomic units) are shown for a simulation of a cell of 32 argon atoms
with a Lennard-Jones
pair potential submitted to uniaxial loading. One of the diagonal
components of the applied stress is increased from zero to 
$15 \times 10^{-5}$ a.u.\ 
during the first 4000 simulation steps and held constant thereafter,
while all the other components are held at zero.
During the first 10000 steps the system is kept in contact with
a heath bath at 10~K. Thereafter we minimize the generalized enthalpy
as described in the text. As the minimization is very fast, the 
horizontal scale is multiplied by a factor of 20 in that region.
During the molecular dynamics the internal stress (wiggly lines)
oscillate around the applied stress (straight lines) as it should.
The minimization makes the internal stress equal to the external stress
within the tolerance of the minimization procedure.
At $\sim 2500$ molecular dynamics steps the system yields in the
way described in the text.
}
\label{fig_stress}
\end{figure}

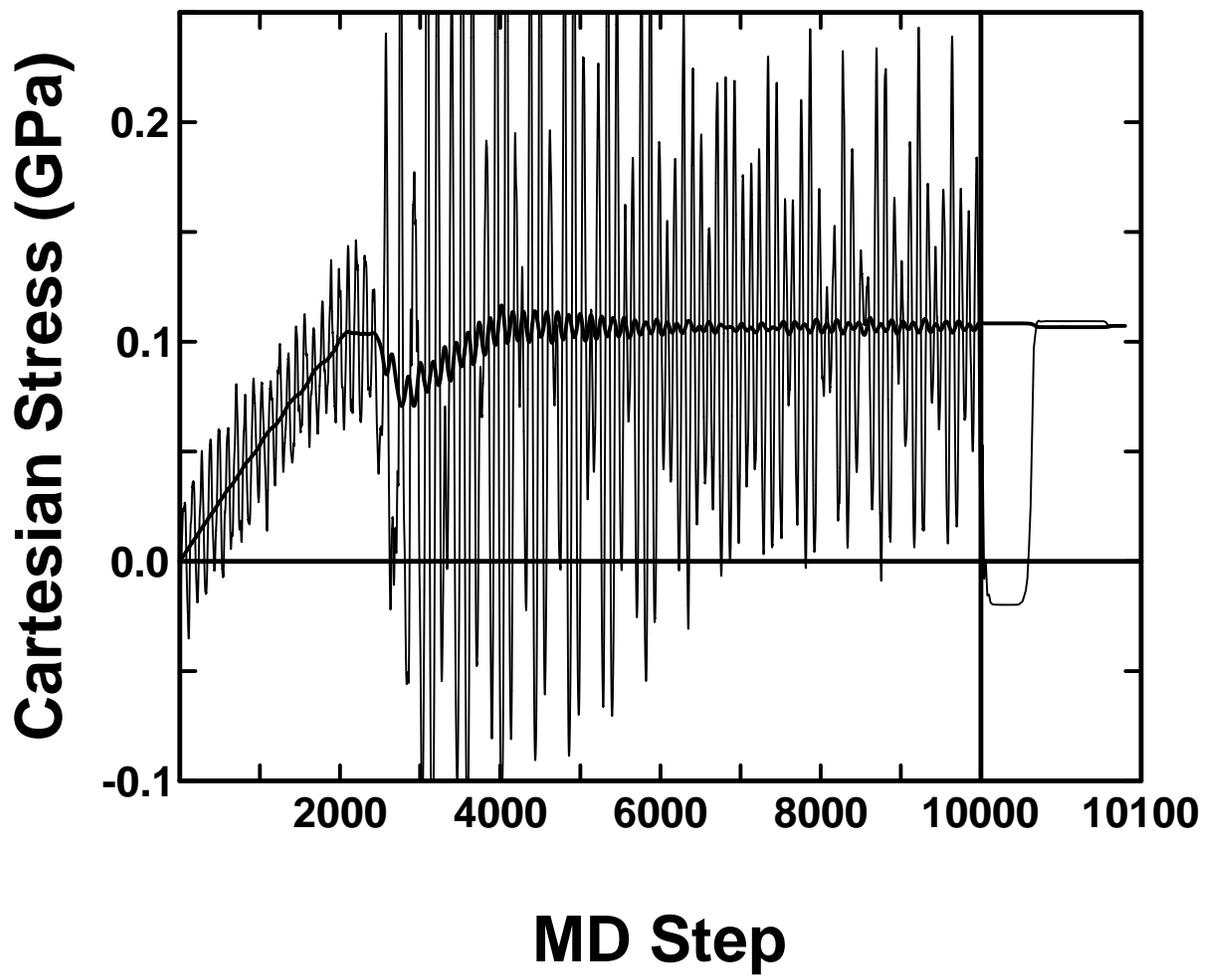
\begin{figure}   
\caption{
The cartesian components of the applied and internal stress tensor along the
direction of compression are shown for the same simulation as in Fig.~2. The
applied stress, which is only indirectly controlled through its contravariant
lattice components, also oscillates. In particular, during the phase transformation
the applied stress drops considerably in response to the decrease in the average
internal stress due to the atomic rearrangement.
}
\label{fig_stress_cart}
\end{figure}

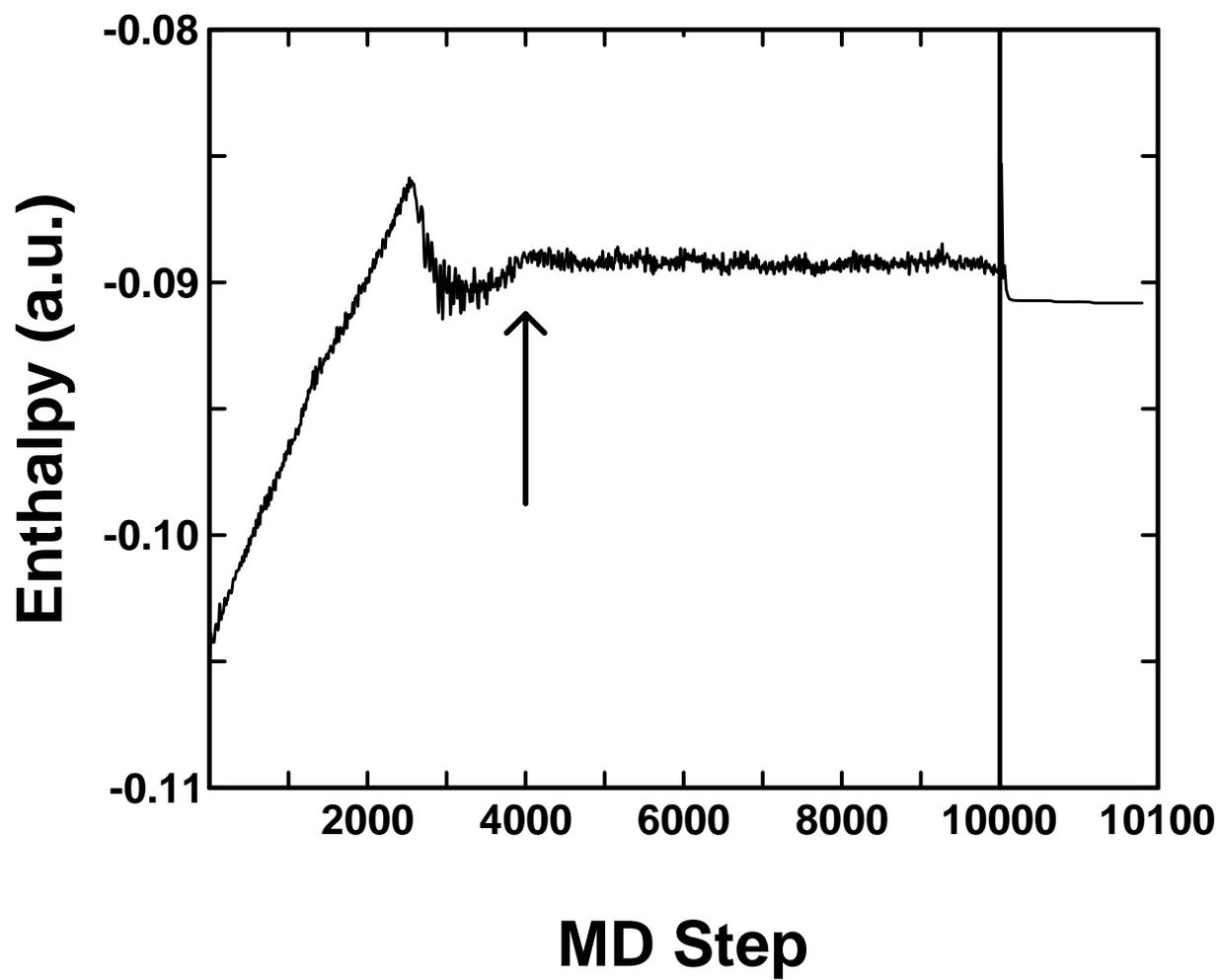
\begin{figure}   
\caption{
The potential component
of the generalized enthalpy (in atomic units) is shown as a function of time
for the same simulation as in Fig.~2.
First we observe an increase of the enthalpy during load due to the
work done on the system by the uniaxial stress. When the system
yields there is a
strong decrease of the potential component of the enthalpy,
even while we continue loading the system, and only near the
end of the loading cycle (indicated by the arrow) do we see 
the enthalpy rising again.
The horizontal scale is again multiplied by a factor of 20 in the
minimization part of the simulation, showing the efficiency of
the procedure of enthalpy minimization in an expanded scale.
}
\label{fig_enth}
\end{figure}

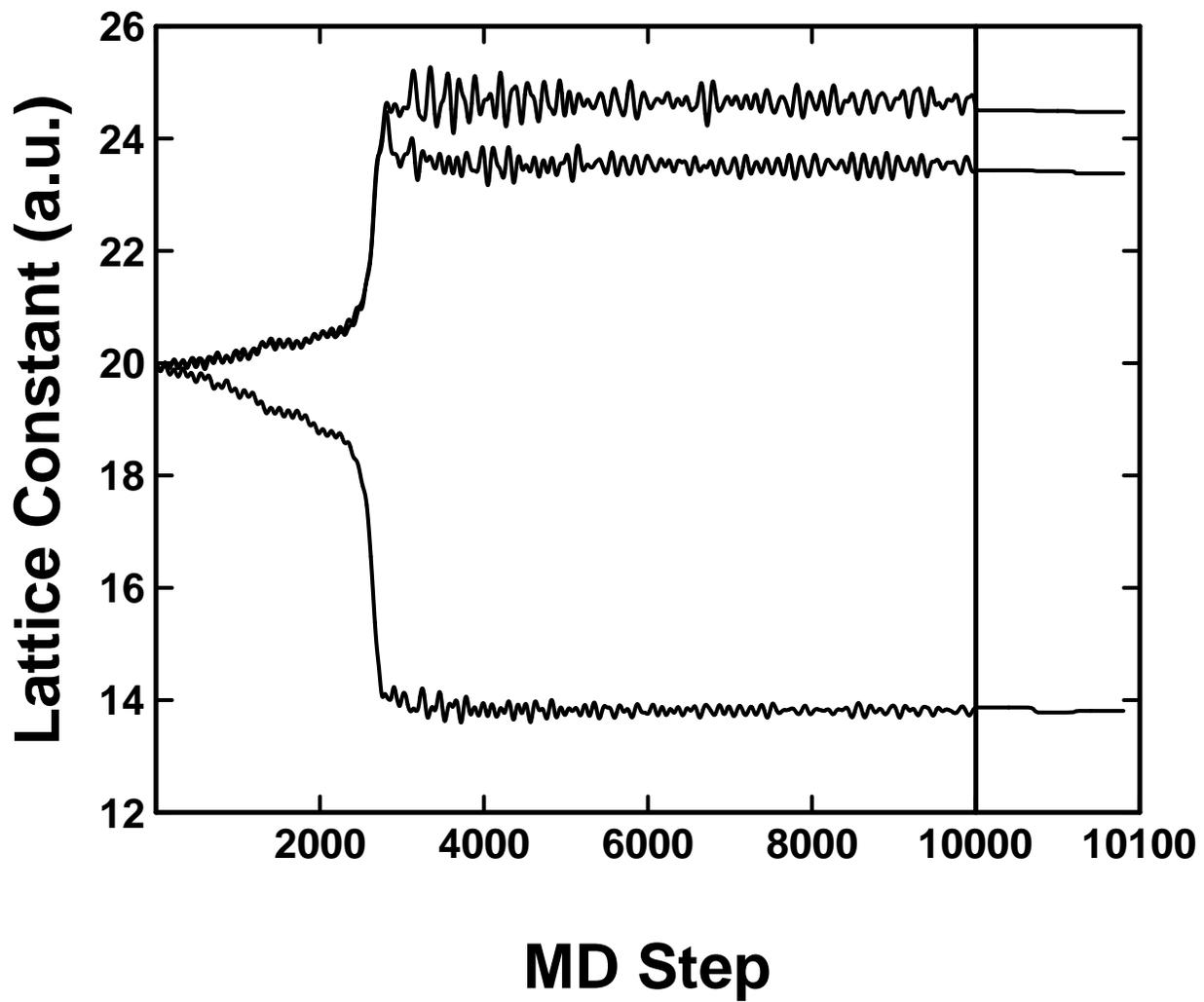
\begin{figure}   
\caption{
The evolution of the three lattice constants (in atomic units) with time is shown
for the same simulation of the previous figures (Figs.~2, ~3 and~4). The strong
structural rearrangement during yielding is clearly seen.
}
\label{fig_latt}
\end{figure}

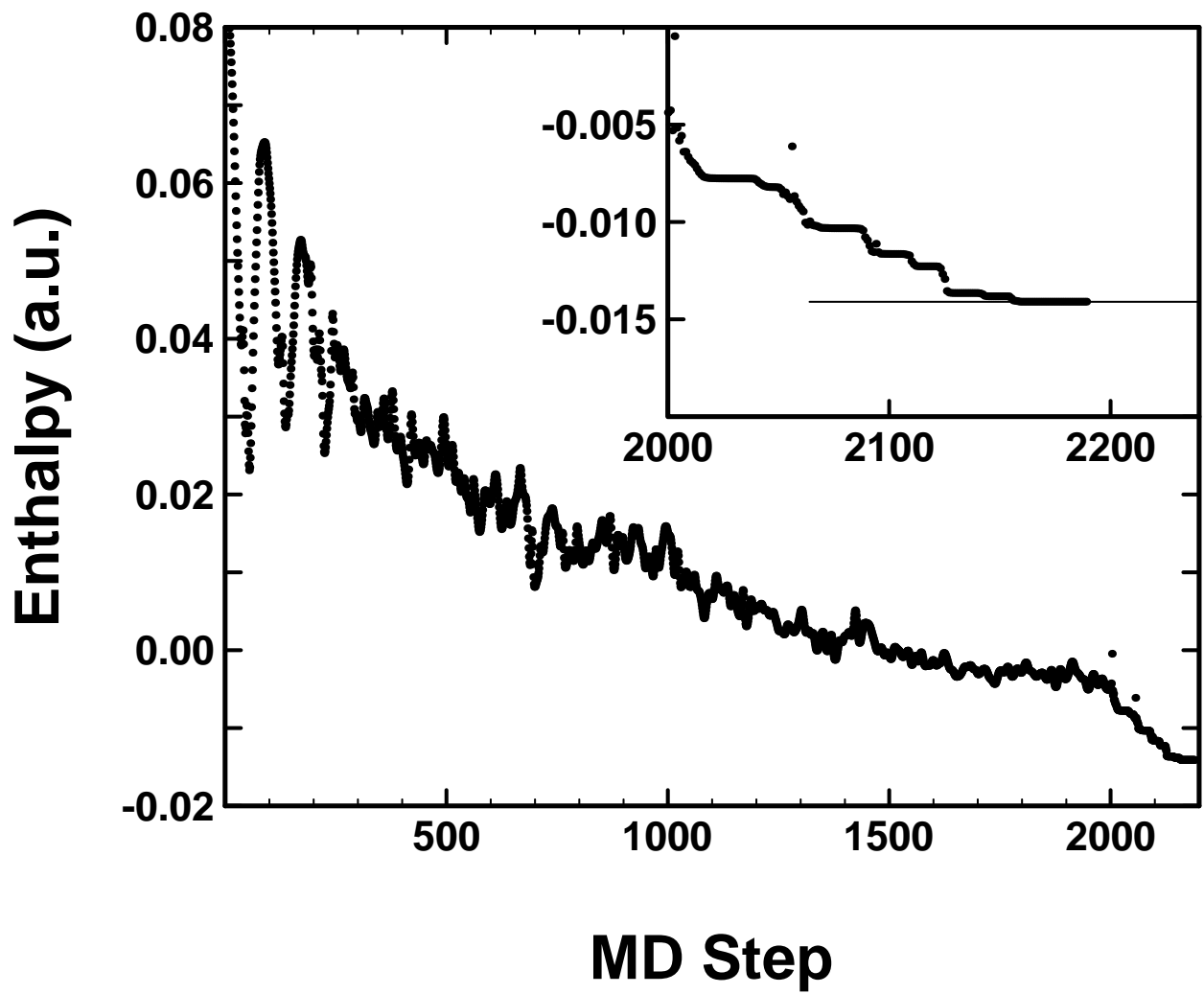
\begin{figure}   
\caption{
The enthalpy (potential part only and in atomic units) of a cell with 16 atoms of Lennard-Jones
argon is shown for a simulation with an
applied pressure of 0.3~GPa.
The simulation starts at conditions quite away from equilibrium, evolves for
2000 steps in contact with a heat bath, and then the enthalpy is minimized.
The inset shows the minimization part of the simulation in an expanded scale.
The horizontal line in the inset is the enthalpy of Lennard-Jones
argon at 0.3~GPa, and that is the value reached by the minimization
procedure. Dots that seem out of place in the minimization
correspond to overshooting steps in the multi-dimensional
minimization procedure.
}
\label{fig_iso}
\end{figure}


\begin{references}
\bibitem{RR_did_it} This equation is in fact equivalent to the second of
Eqs. (2.26) in Ref.\cite{RR}.
\bibitem{footnote1}  When the infinitesimal deformation is not rotation-free,
$h=\left( 1+ \epsilon \right) h_0$ can be written as 
$h\simeq \left( 1+ \epsilon _S \right) \left( 1+ \epsilon _A \right) h_0$,
where $1 + \epsilon _A$ describes a pure infinitesimal rotation and
$1 + \epsilon _S$ describes a rotation-free infinitesimal deformation ($\epsilon _S$
and $\epsilon _A$ are respectively the symmetrized and
anti-symmetrized $\epsilon$).
The relation between $\epsilon$ and the Lagrangian strain
$\varepsilon$ is $\varepsilon = { 1 \over 2 } \left( \epsilon + \epsilon ^T +
\epsilon ^T \epsilon \right)$. For small deformations, $\varepsilon$ is
the symmetrical version of $\epsilon$.
\end{references}
\end{document}